\documentclass[12pt]{iopart}

\usepackage{epsfig}
\newenvironment{narrow}[2]{
   \begin{list}{}{
      \setlength{\topsep}{0pt}
      \setlength{\leftmargin}{#1}
      \setlength{\rightmargin}{#2}
      \setlength{\listparindent}{\parindent}
      \setlength{\itemindent}{\parindent}
      \setlength{\parsep}{\parskip}}
\item[]}{\end{list}}

\begin{document}

\title[Dielectron Production in $^{12}$C + $^{12}$C Collisions at \mbox{1 GeV/u}]{Dielectron Production in $^{12}$C + $^{12}$C Collisions at \mbox{1 GeV/u} and the Solution to the DLS Puzzle}

\begin{noindent}
\author{Y. C. Pachmayer (for the HADES collaboration)}
\address{\mbox{Institut~f\"{u}r~Kernphysik,~Johann~Wolfgang~Goethe-Universit\"{a}t,}~\mbox{Frankfurt,~Germany}}
\ead{Y.Pachmayer@gsi.de}
\end{noindent}

\begin{center}
\vspace{5 mm}
{G.~Agakishiev$^{8}$, C.~Agodi$^{1}$, A.~Balanda$^{3,e}$, G.~Bellia$^{1,a}$,
D.~Belver$^{15}$, A.~Belyaev$^{6}$, A.~Blanco$^{2}$, M.~B\"{o}hmer$^{11}$, J.~L.~Boyard$^{13}$,
P.~Braun-Munzinger$^{4}$
% ...
}\newline

(HADES collaboration)
\end{center}

\begin{tabular}{rl}

$^{1}$ &Instituto Nazionale di Fisica Nucleare - Laboratori Nazionali del Sud, \\ & 95125 Catania, Italy\\
$^{2}$ &LIP-Laborat\'{o}rio de Instrumenta\c{c}\~{a}o e F\'{\i}sica Experimental de Part\'{\i}culas , \\ & 3004-516 Coimbra, Portugal\\
$^{3}$&Smoluchowski Institute of Physics, Jagiellonian University of Cracow, \\ & 30-059~Krak\'{o}w, Poland\\

\end{tabular} \newline

\begin{tabular}{rl}
$^{4}$&Gesellschaft f\"{u}r Schwerionenforschung mbH, 64291~Darmstadt, Germany\\
$^{5}$&Institut f\"{u}r Strahlenphysik, Forschungszentrum Dresden-Rossendorf, \\ & 01314~Dresden, Germany\\
$^{6}$&Joint Institute of Nuclear Research, 141980~Dubna, Russia\\
$^{7}$&Institut f\"{u}r Kernphysik, Johann Wolfgang Goethe-Universit\"{a}t, \\ &  60438 ~Frankfurt, Germany\\
$^{8}$&II.Physikalisches Institut, Justus Liebig Universit\"{a}t Giessen, 35392~Giessen, \\ & Germany\\
$^{9}$&Istituto Nazionale di Fisica Nucleare, Sezione di Milano, 20133~Milano, Italy\\
$^{10}$&Institute for Nuclear Research, Russian Academy of Science, 117312~Moscow, \\ & Russia\\
$^{11}$&Physik Department E12, Technische Universit\"{a}t M\"{u}nchen, 85748~M\"{u}nchen, \\ & Germany\\
$^{12}$&Department of Physics, University of Cyprus, 1678~Nicosia, Cyprus\\
$^{13}$&Institut de Physique Nucl\'{e}aire (UMR 8608), CNRS/IN2P3 - Universit\'{e} \\ & Paris Sud, F-91406~Orsay Cedex, France\\
$^{14}$&Nuclear Physics Institute, Academy of Sciences of Czech Republic, 25068~Rez, \\ & Czech Republic\\
$^{15}$&Departamento de F\'{\i}sica de Part\'{\i}culas, University of Santiago de Compostela, \\ &  15782~Santiago de Compostela, Spain\\
$^{16}$&Instituto de F\'{\i}sica Corpuscular, Universidad de Valencia-CSIC, \\ & 46971~Valencia,  Spain\\
\\
$^{a}$& also at Dipartimento di Fisica e Astronomia, Universit\`{a} di Catania, \\  & 95125~Catania,  Italy\\
$^{b}$& also at ISEC Coimbra, ~Coimbra, Portugal\\
$^{c}$& also at Technische Universit\"{a}t Dresden, 01062~Dresden, Germany\\
$^{d}$& also at Dipartimento di Fisica, Universit\`{a} di Milano, 20133~Milano, Italy\\
$^{e}$& also at Panstwowa Wyzsza Szkola Zawodowa , 33-300~Nowy Sacz, Poland\\
\end{tabular} \newline \newline

\begin{abstract}
The production of $e^{+}e^{-}$ pairs in $^{12}$C + $^{12}$C collisions at \mbox{1 GeV/u} was investigated with the HADES experiment at GSI, Darmstadt. In the invariant-mass region $ 0.15\: GeV/c^{2} \leq  M_{ee} \leq 0.5\: GeV/c^{2}$ the measured pair yield shows a strong excess above the contribution expected from hadron decays after freeze-out. The data are in good agreement with the results of the former DLS experiment for the same system and energy.
\end{abstract}


\section{Introduction}
Dilepton spectra taken in heavy-ion collisions exhibit an enhancement of yield, independent of the bombarding-energy,  compared to the superposition of free hadronic decays in the invariant-mass region $ 0.2\: GeV/c^{2} \leq  M_{ee} \leq 0.6\: GeV/c^{2}$. While the dilepton enhancement found at SPS energies has been related to modifications of the $\rho$-meson spectral function in the hadronic medium \cite{paper3}, the large pair yields observed by DLS \cite{paper1} in C+C and Ca+Ca collisions at 1 GeV/u remain to be explained \mbox{satisfactorily \cite{paper4}}.
The High-Acceptance DiElectron Spectrometer HADES \cite{paper2} at GSI, Darmstadt, is presently the only dilepton spectrometer operational in the SIS energy regime of \mbox{1-2 GeV/u}, succeeding the DLS experiment. First results obtained in C+C collisions at 2 GeV/u confirmed the general observation
of an enhanced pair yield above the contribution expected from hadron decays after freeze-out \cite{paper5}. In this note we report on a measurement of inclusive electron-pair emission from $^{12}$C+$^{12}$C collisions at 1 GeV/u. Taken together, both results allow to discuss the beam energy dependence of the pair yield. Furthermore a direct comparison with the DLS results \cite{paper1} has become possible, thus addressing (and solving experimentally) the long-standing "DLS puzzle".

\section{Dielectron Production at \mbox{1 GeV/u} kinetic beam energy}
\begin{figure}
\begin{narrow}{-.5in}{0in}
  \centering
       \centering
     \begin{minipage}[c]{0.45\linewidth}
 \centering
        \caption[]{
  \small Dielectron yield (corrected for inefficiencies)
in the HADES acceptance. In a) the measured yield is
compared to a cocktail calculated from sources assuming vacuum
properties only.
The cocktail is divided into contributions from long-lived mesons
$\pi^{0}, \eta, \omega$ (full line, cocktail A)
and contributions from short-lived resonances $\rho$, $\Delta$; the sum of all defines cocktail B.
b) shows the experimental yield divided by cocktail A
for 1 GeV/u (full symbols) and 2 GeV/u (open symbols) data.
In addition, the ratio of cocktail B and A for 1 GeV/u data
is indicated (dashed line).
  }\label{Fig1}
  \end{minipage}
     \hspace{0.02\linewidth}
     \begin{minipage}[c]{0.45\linewidth}
       \includegraphics[width=\linewidth]{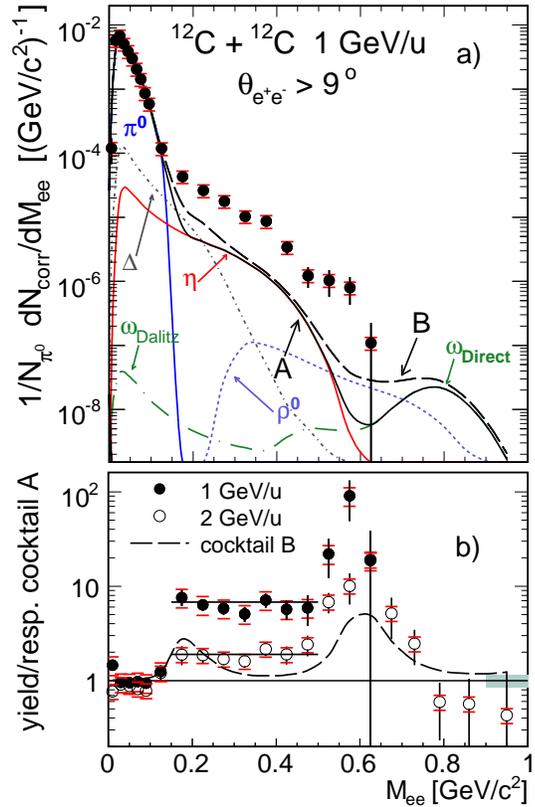}
    \end{minipage}

\end{narrow}
\end{figure}

The dielectron yield measured in HADES in $^{12}$C+$^{12}$C collisions at 1 GeV/u was
corrected for detection and reconstruction inefficiencies as described
in \cite{paper6}. Fig.~\ref{Fig1} shows the resulting $e^+e^-$
invariant-mass distribution of true pairs normalized to the average
number of charged pions $N_{\pi^0} = 1/2 (N_{\pi^+} + N_{\pi^-})$,
as measured also in HADES and extrapolated to the full solid angle.  The
obtained pion multiplicity per participant nucleon, i.e.
$M_{\pi}/A_{part} = 0.061 \pm 0.009$, agrees well with previous
measurements of charged and neutral pions \cite{paper7}.
The quoted error of $15\%$ is dominated by systematic uncertainties
in the pion efficiency correction and the extrapolation procedure.
In ad\-dition to this overall normalization error, uncertainties caused
by the electron-efficiency correction and by the subtraction of the
combinatorial background add up quadratically to point-to-point
systematic errors in the pair yield of $22\%$ (also shown in Fig.~\ref{Fig1}).

\begin{figure}
\begin{narrow}{-.5in}{0in}
  \centering
       \centering
     \begin{minipage}[c]{0.45\linewidth}
 \centering
        \caption[]{
  \small Dielectron yield as a function of pair $P_{\bot}^{ee}$
  for the invariant-mass region $ 0.15\: GeV/c^{2} \leq  M_{ee} \leq 0.5\: GeV/c^{2}$.
  Line codes as in Fig.~\ref{Fig1}.
   }\label{Fig2}
  \end{minipage}
     \hspace{0.02\linewidth}
     \begin{minipage}[c]{0.45\linewidth}
       \includegraphics[width=\linewidth]{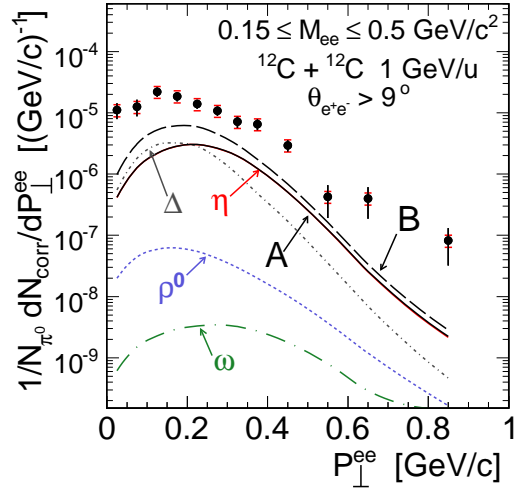}
    \end{minipage}
\end{narrow}
\end{figure}

\noindent We compare our data with a pair cocktail (cocktail A), which was calculated from $\pi^0$, $\eta$ and $\omega$ meson decays to represent the radiation from long-lived mesons (decaying mostly outside the fireball).
The $\pi^0$ and $\eta$ Dalitz yields are constrained by published data \cite{paper7}; for the production rate of the $\omega$ meson we apply $m_T$-scaling \cite{mt_scaling}.
The cocktail calculation was performed with the PLUTO generator
assuming anisotropic meson emission from a Boltzmann-like thermal source
(for details cf. \cite{paper6}). While experimental data and \mbox{cocktail A} agree
in the $\pi^0$ Dalitz region, the cocktail strongly underestimates the measured pair
yield for $M_{ee}>0.15$ $GeV/c^{2}$.  This is not surprising, since one expects additional contributions
from short-lived resonances, e.g. $\Delta(1232)$ and $\rho$. For the $\Delta$, we assumed
that its contribution scales with the $\pi^0$ yield at freeze-out
and that the $\Delta\rightarrow Ne^+e^-$ differential
decay rates of \cite{paper4} are applicable. We modeled the broad resonance $\rho$
as a Breit-Wigner shape, with mass-dependent
width $\Gamma(M)=\Gamma_0/M^3$ ($\Gamma_0=0.15$ GeV) \cite{paper4},
additionally modified by $m_T$-scaling accounting for the strongly
reduced phase space at low beam energies.  The resulting cocktail B is
shown in Fig.~\ref{Fig1} (long-dashed line).
Adding these short-lived contributions increases the simulated yield
above $0.15$ $GeV/c^{2}$, but obviously our second calculation also falls short
of reproducing the data.  More sophisticated calculations, e.g. based
on transport models, are clearly needed. \newline
\noindent For better visualization of the character of the excess yield, the ratio of data
and cocktail A is shown in Fig.~\ref{Fig1}(b).
This ratio is basically unity at low masses, where $\pi^0$ Dalitz pairs
dominate, but above $M_{ee} = 0.15$ $GeV/c^{2}$ it is large, indicating the onset
of processes not accounted for by our cocktail A.
Fig.~\ref{Fig1}(b) also shows the corresponding ratio
observed at 2 GeV/u \cite{paper5}.
It is evident that at 1 GeV/u the overshoot of the data is much
stronger than at 2 GeV/u. A detailed analysis shows that the beam energy dependence of the excess yield above the
known $\eta$ contribution \cite{paper7}, integrated over the
$ 0.15\: GeV/c^{2} \leq  M_{ee} \leq 0.5\: GeV/c^{2}$ mass range, scales
like $\pi$ production \cite{paper6}. \newline \noindent
Fig.~\ref{Fig2} shows the pair transverse momentum ($P_{\bot}^{ee}$)
distribution for the excess region $ 0.15\: GeV/c^{2} \leq  M_{ee} \leq 0.5\: GeV/c^{2}$.
Note that in comparison to cocktail A the excess factor is constant at high $P_{\bot}^{ee}$, but increases steeply towards lower $P_{\bot}^{ee}$.

\begin{figure}
\begin{narrow}{-.5in}{0in}
  \centering
       \centering
     \begin{minipage}[c]{0.45\linewidth}
 \centering
        \caption[]{
  \small Direct comparison of the dielectron cross sections
   measured in the reaction
$^{12}$C+$^{12}$C at 1 GeV/u by HADES (full triangles)
and at 1.04 GeV/u by DLS \cite{paper1} (empty triangles).
   The invariant-mass distributions are compared within the DLS acceptance.
   Statistical and systematic errors are shown.  Overall normalization
   errors (not shown) are $20\%$ for the HADES and $30\%$ for the DLS
   data points.
   }\label{Fig3}
  \end{minipage}
     \hspace{0.02\linewidth}
     \begin{minipage}[c]{0.45\linewidth}
       \includegraphics[width=\linewidth]{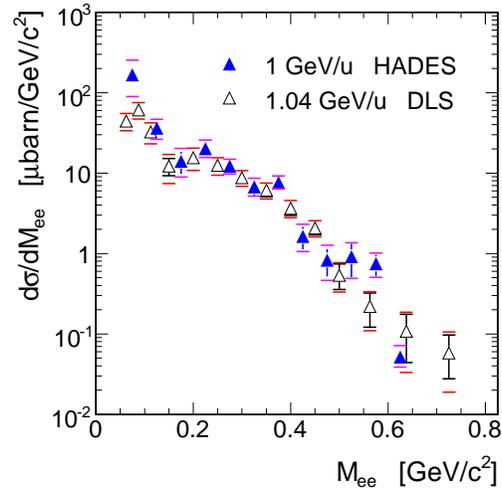}
    \end{minipage}
\end{narrow}
\end{figure}

\noindent A direct comparison of HADES and DLS \cite{paper6} results can be achieved by a mapping of the measured HADES pair yield onto the DLS acceptance, defined in the 3-d space spanned by the pair variables
$M_{ee}, P_{\perp}^{ee}$ and the rapidity $Y_{ee}$.  Although the \mbox{acceptances of} both ap\-paratuses
do not fully overlap for low-mass, low-$P_{\perp}^{ee}$ pairs, in the excess region,
the HADES coverage is larger and almost fully contains the DLS acceptance.
Transforming the multiplicities measured by HADES to cross sections,
this mapping allows for an almost model-independent comparison of the two
data sets (for details cf. \cite{paper6}). In Fig. \ref{Fig3} the HADES-mapped invariant-mass
distribution is shown together with the DLS \mbox{result \cite{paper1}.}
It is apparent that,
within statistical and systematic uncertainties, both
measurements are in agreement, in particular in the region
of the excess yield.  The same conclusion is obtained from the comparison of
the $P_{\perp}^{ee}$ distributions \cite{paper6}.

\section{Conclusion}
The agreement of the present data with the -- for a long time disputed -- results \mbox{in \cite{paper1}} solves the "DLS puzzle" experimentally. It poses, however,
again the question of
the origin of the pair excess. In this
context, studies of the elementary re\-actions $p+p$ and $d+p$ are important steps.
Indeed, recent calculations within a One Boson Exchange (OBE)
model \cite{kaptari} suggest significantly larger than heretofore assumed
contributions from $p-p$ and, mostly, $p-n$ quasi-elas\-tic bremsstrahlung.
Moreover, transport calculations done with the Hadron String Dynamics (HSD)
model \cite{HSD07} using a parametrization of bremsstrahlung inspired
by the new OBE result \cite{kaptari} seem to match both the HADES and the DLS
$^{12}$C+$^{12}$C data.  In this situation it is evident that the direct confrontation of the
OBE model calculations with $p+p$ and $d+p$ dilepton data measured with HADES \cite{hades3} is
mandatory to reach firm conclusions on the origin of dileptons at
SIS energies.

\end{document}